\documentclass{elsart}
\usepackage{natbib}
\usepackage{graphicx}
\begin{document}
\begin{frontmatter}

\title{New analysis in the field of open cluster Collinder 223}
{\author Tadross, A. L.$^\star$}\\
\footnotetext[1]{E-mail address: altadross@mailer.scu.eun.eg (A.L.
Tadross)}

\address{National Research Institute of Astronomy and Geophysics,
11421- \\ Helwan, Cairo, Egypt. (Tel.:+20-2-5560-645,
fax:+20-2-5548-020)}

\begin{abstract}
The present study of the open cluster Collinder 223 (Cr 223) has
been mainly depended on the photoelectric data of Clari$\acute{a}$
\& Lapasset \cite{CL91}; hereafter CL91. This data of CL91 has
been used with the cluster's image of AAO-DSS$^{1}$ in order to
re-investigate and improve the main parameters of Cr 223. Stellar
count has been achieved to determine the stellar density, the
cluster's center and the cluster's diameter. In addition, the
luminosity function, mass function, and the total mass of the
cluster have been estimated.
\end{abstract}

\end{frontmatter}

\footnotetext[1]{AAO: Anglo-Australian Observatory - DSS:
Digitized Sky Surveys; taken from "SIMBAD"
(http://simbad.u-strasbg.fr)}

\section*{1. Introduction}

The open cluster Cr 223 (C1028-595) situated in the Carina spiral
feature of the southern Milky Way with 2000.0 coordinates $
\alpha=10^{h} 30.5^{m}, \delta=-59^{\circ} 49^{'}, \ell=
286.2^{\circ}, b= -1.9^{\circ}$. Trumpler \cite{TR30} classified
this cluster as a detached cluster with little central
concentration and a medium range in the brightness of the stars
(II-2p system). Collinder \cite{CO31} described this cluster as a
system of 38 stars at a distance of 2630 pc distributed over a
field of 8 arcmin.

110 stars in the cluster region have been studied by CL91 used the
broadband {\it UBV} system and moreover seven probable red giants
have been observed used the intermediate-band of {\it DDO} system.
CL91 proved that these stars represented a genuine open cluster of
only {\it B-type} stars that extends over a {\it V-range} of about
11-14 mag. The identification chart of Cr 223 has been taken from
Hogg \cite{HO65} and presented in Figure 1 of CL91. It is defined
a radius of about 7.5 arcmin from the apparent center of the chart
(near the star no. 77). CL91 described their work as a preliminary
photometric study of Cr 223. Therefore some of the present
parameters e.g., the luminosity and mass functions are preliminary
also. So that more deep studies (CCD observations) of the cluster
Cr 223 would be needed for more precisely determination of these
parameters.

In the present work, reddening, distances, cluster's center,
cluster's diameter, stellar density, age, membership analysis,
luminosity function, mass function, and the total mass of the
cluster have been estimated.

\section*{2. HR-Diagrams}

The main photometrical parameters of the cluster (membership,
reddening and distance) are depending mainly on HR-diagrams of
that cluster, i.e. color-color (CC) diagram [{\it (U-B)-(B-V)}],
short and long wavelengths color-magnitude diagrams CMDs [{\it
V-(U-B) \& V-(B-V)}].

\subsection*{2.1. Membership analysis}
The uncertainty about membership increases for faint stars those
are merging with the crowded galactic field. To discriminate
between cluster members and field stars, the criteria of
Clari$\acute{a}$ \& Lapasset \cite{CL86} has been applied
and the members have been chosen to verify the two conditions: \\
(1) its location in the two CMDs must correspond to the same
evolutionary stage in the cluster, and (2) its location in the
(CC) diagram must be close to the cluster main sequence, the
maximum departure accepted being about 0.1 mag.

CL91 classified 46 of 110 stars as members including two blue
stragglers (nos. 35 \& 80); and one red giant (no. 75); and three
probable members (nos. 30, 81 \& 95). They classified about 60
stars as non members. For more details about the field stars see
CL91.

Applying the above criteria after separating the giant stars and
field ones, it is found that three stars could be added to the
cluster Cr 223 as members (nos. 23, 31 \& 32). Although these
three new members are quite away from the cluster center, they lie
inside the cluster's extended diameter (area) that estimated in
section 6 of this work. On the other hand, their ranges of
distance and reddening are found in agreement with those of the
cluster's members.

\subsection*{2.2. Reddening and distance}
The presence of interstellar matter throughout the spiral arms of
the Galaxy makes the distance determination of the galactic
clusters somewhat difficult where the starlight is frequently
absorbed and scattered.

CL91 assumed that the cluster Cr 223 has some differential
reddening, and for that reason they used equations (2) and (8) of
Garcia, Clari$\acute{a}$ \& Levato \cite{GCL88} to dereddened
stars individually with standard deviation of 0.03 mag.

In the present work, reddening (which is the importance parameter
in deriving distance) has been estimated simultaneously with the
apparent distance modulus with a standard deviation of 0.015 mag.
These two parameters are estimated by fitting the standard
zero-age main sequence {\it ZAMS} of Schmidt-Kaler \cite{SCHM82}
to the lower envelope of the points in both CMDs of the cluster.
Many fittings have been applied to reach the best corresponding
values of reddening at the same distance moduli. The evolved stars
have been excluded from the fitting and the interstellar
absorption law has been applied. For each fit, the calculated
value of E(U-B) has been computed using the relation:
\begin{center}
{\it E(U-B)$_{cal.}$ = 0.72 * E(B-V) + 0.05 * E$^{2}$(B-V)}
\end{center}
At the minimum difference between the observed and calculated
values [$\triangle$ E(U-B)$_{Cal. - Obs.}$ $\approx$ 0.0 mag],
reddening values and the related distance modulus have been found
to be E(U-B)= 0.18 $\pm$ 0.015 mag, E(B-V)= 0.25 $\pm$ 0.015 mag,
and V-M$_{v}$ =13.0 $\pm$ 0.15 mag. The resulting total visual
absorption is then A$_{v}$ =0.75 mag where the value of the ratio
A$_{v}$/E(B-V) is taken to be 3.0 as assuming by Garcia et al.
\cite{GCL88}.

Comparing the observed (CC) diagram of the cluster Cr 223 with the
standard one of Schmidt-Kaler \cite{SCHM82}, the intrinsic values
of the color indices are estimated for each star as follows: A
line parallel to the reddening line has been drawn for each star
and the intersection of this line with the {\it ZAMS}-curve gives
the intrinsic color indices [{\it (B-V)$_{o}$ \& (U-B)$_{o}$}]
assuming that the star lies on the main sequence. The slope of the
reddening line has been taken to be 0.72 as given by Johnson \&
Morgan \cite{JOMO53}. The stars which are lying below the kink of
the {\it ZAMS}-curve are ambiguous stars and may have two or three
possible values. For such stars, the best reading that is
consistent with the cluster distance has been taken. The intrinsic
visual magnitude [{\it V$_{o}$}] for each star is determined from
the individual reddening estimation.

The two free CMDs of the cluster have been constructed for
cluster's members as shown in Fig 1. The solid curved lines
represented the standard {\it ZAMS}-curves of Schmidt-Kaler
\cite{SCHM82} fitted to the lower envelope of the points in the
two CMDs. The true distance modulus then is {\it
(V-M$_{v}$)$_{o}$} =12.25 $\pm$ 0.15 mag , which equals a distance
of 2820 $\pm$ 190 pc. This distance is found to be in agreement
with what obtained by CL91.

The distances of the cluster from the galactic plane ({\it Z}),
from the galactic center ({\it Rgc}), and the projected distances
from the Sun on the galactic plane ({\it X \& Y}) have been
calculated to be -81 pc, 8.2 kpc, -2.7 kpc and 0.8 kpc
respectively.

\section*{3. Stellar Density}

Counting stars of the cluster through the interstellar medium,
particularly in southern sky allow us to define the obscuration
clouds before the face of the cluster. The whole studied area of
the cluster on AAO-DSS image can be seen in Fig 2.

Applying Wallenquist's \cite{WALL75} method using the AAO-DSS
image of Cr 223, an area of about 3024 arcmin$^{2}$ has been
covered and more than 2000 stars have been counted. For more
details about similar task see Tadross et al. \cite{TAD02}.

A contour map of a grid of 1400 density points has been generated
for the cluster region. The stellar density at each grid point has
been calculated for the whole area, which appears that it
concentrates in south-west direction. It may be a resultant of
intrinsic spatial distribution in this area, or caused by some
clouds that lie on the face of the cluster. If so, these clouds
should affect the stellar densities of the areas behind them and
then the concentration of the stellar density in south-west
direction is related to the lower values of obscuration as shown
in the contour map of Fig 3.

\section*{4. Age}

CL91 used the bluest color indices to estimate the age of the
cluster Cr 223. They found that it has an age of 3.6*10$^{7}$ yr,
which makes the cluster belonging to the IC 4665 age group of
Mermilliod \cite{MER81}.

In the present work, the age of the cluster has been estimated
applying the isochronous curves of Meynet et al. \cite{MEY93} on
the free CMD of Cr 223 [{\it $V_{o}-(B-V)_{o}$}] as shown in Fig
4. The evolved sequence of the cluster implies that it is in
agreement with an age of about 10$^{8}$ yr, or younger.

\section*{5. Cluster's center}

The position of the cluster's center has been determined by
counting the stars in two orthogonal rectangular strips using
AAO-DSS image of the cluster. The strips were aligned with
$\alpha$ and $\delta$ directions and divided into suitable bins
along their lengths. The densities of the bins are plotted against
the central positions of the strips along $\alpha$ and $\delta$
respectively; see Tadross et al. \cite{TAD02}. The center of
symmetry about the peaks was taken to be the cluster's center as
shown in Fig 5. The new center has found to be shifted from CL91's
center by 1 arcmin in the northwest direction, i.e. it lies at
$\alpha = 10^{h}:32^{m}:16^{s}$ and $\delta =
-60^{o}:1^{'}:12^{"}$, see Fig 2.

\section*{6. Angular and Linear Diameter}

Knowing the plate scale of AAO-DSS image we can easily see how far
the cluster extends and hence estimate the angular diameter and
consequently the linear diameter as well. Comparing the chart of
Hogg \cite{HO65} with the AAO-DSS image of Cr 223, it is found
that the members of CL91 are distributed in a circle of about 15
arcmin in diameter. On the other hand, an estimation of the
angular diameter can be obtained by examining the radial density
distribution of the cluster's stars. For that purpose the
projected stellar density of Cr 223 has been a determined counting
star in 16-concentric rings from the new center of the cluster.
The real stellar density in the cluster's area has been
constructed in a histogram as shown in Fig 6. A radius of about 9
arcmin has been obtained which equivalent to 7.4 pc.

\section*{7. Luminosity Function}

Two histograms have been constructed for the cluster's field and
members where the numbers of stars have been counted at intervals
of 0.5 mag on {\it V}-scale. Half magnitude intervals have been
selected to include a reasonable number of stars per magnitude bin
and for the best possible statistics in the luminosity function.

The luminosity of each member star of the cluster has been
calculated from its visual magnitude. The total luminosity of Cr
223 is calculated to be -4.4 mag summing up the luminosity of
every member star it has. However, to show what is called the net
observed luminosity function of the cluster, the histogram of the
field stars is subtracted from the histogram of the cluster's
members as shown in Fig 7. The absolute magnitude scale appears on
the upper axis of that figure with the limiting magnitude:
\begin{center}
-2.5 $<$ {\it M$_{v}<$} 1.5 mag, or 1.3 $<$ Log
{\it(L/L$_{\odot}$)} $<$ 2.9
\end{center} where Log {\it (L/L$_{\odot}$)} = $0.4*(4.79$ - {\it
M$_{v}$)} and L \& L$_{\odot}$ are the luminosity of the star and
the Sun respectively. The high peak lies at {\it M$_{v}$} = 0.0
mag, or at Log {\it (L/L$_{\odot}$)} $\approx$ 2.0 .

\section*{8. Mass Function and total mass}
The mass function of the cluster Cr 223 has been estimated using
the theoretical evolutionary tracks and their isochronous of
different ages of VandenBergh \cite{VAN85}. The masses of the
cluster's members have been estimated applying the polynomial
equation that has been developed from the isochronous data at the
metallicity factor z=0.0169 and {\it age} $<$ 2.5*10$^{8}$ yr.

The mass function histogram has been constructed dividing the mass
scale into suitable bins and counting the number of stars at each
bin. The high peak lies at 3.75 {\it M$_{\odot}$} as shown in Fig
8. On this concept, summing up the stars in each bin weighted by
the mean mass of that bin yields the total mass of the cluster
that found to be about 190 {\it M$_{\odot}$}.

\section*{9. Conclusions}
It is necessary to state that the photometric study of this
cluster was depended mainly on the previous preliminary work of
CL91 that, of course, has limited number of stars. For that some
estimated parameters of the present work, e.g. luminosity and mass
functions are also preliminary ones and need more deep study for
more precisely determinations. On the other hand, the image of the
cluster Cr 223 in the {\it AAO-DSS} system shows that it needs
more extensive CCD observations to enrich the faint members and
consequently fills the lower part of the main sequence of the
cluster. For comparing the present results with the previous ones
of CL91, see table 1.

{\bf Acknowledgement}

I would like to offer my appreciation to the teamwork of the
Digitized Sky Survey {\it DDS} and the Anglo-Australian
Observatory {\it AAO} for providing such very useful images, which
serve that kind of work $^{2}$.

\footnotetext[2]{http://archive.stsci.edu/dss/acknowledging.html
http://archive.stsci.edu/dss/copyright.html}

\newpage
\begin{figure}
\centerline{\includegraphics[width=10cm]{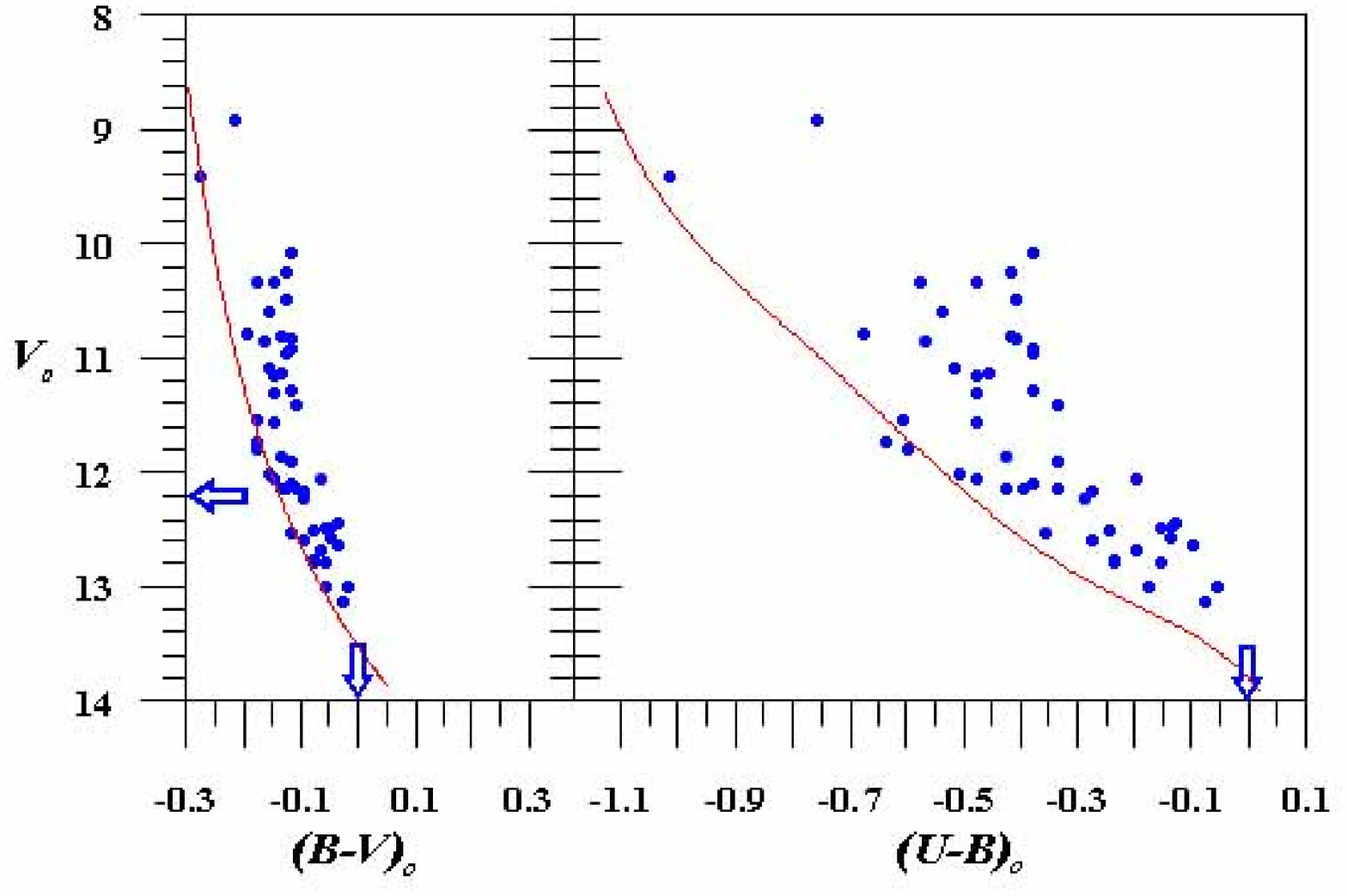}} \caption{The
free CM diagrams of the cluster's members fitted with {\it ZAMS}
of Schmidt-Kaler (1982). The arrows show the locations of {\it
M$_{v}$}=0.0 mag, $(B-V)_{o}$=0.0 mag \& $(U-B)_{o}$=0.0 mag.}
\end{figure}

\begin{figure}
\centerline{\includegraphics[width=10cm]{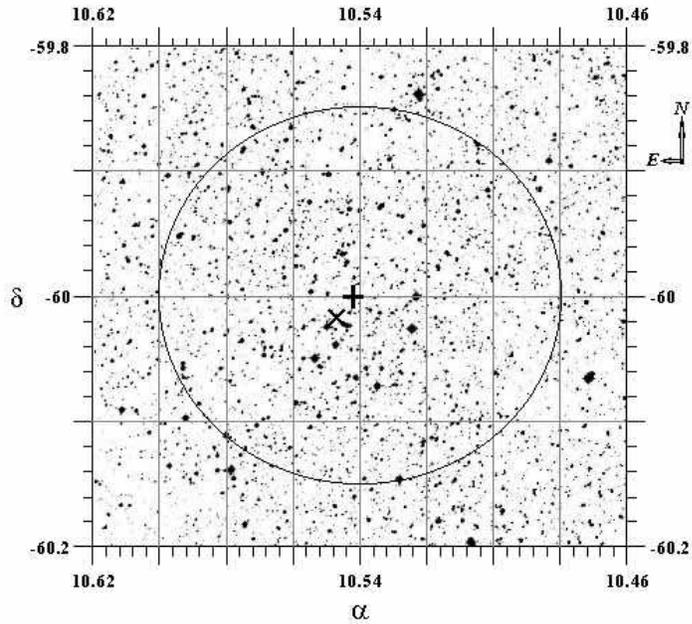}}
\caption{AAO-DSS image of the cluster Cr 223. "$\times$" refers to
the apparent center of the cluster as defined by CL91, "$+$"
refers to the new center of the present work, and the circle
defines the whole studied area of the cluster.}
\end{figure}

\begin{figure}
\centerline{\includegraphics[width=10cm]{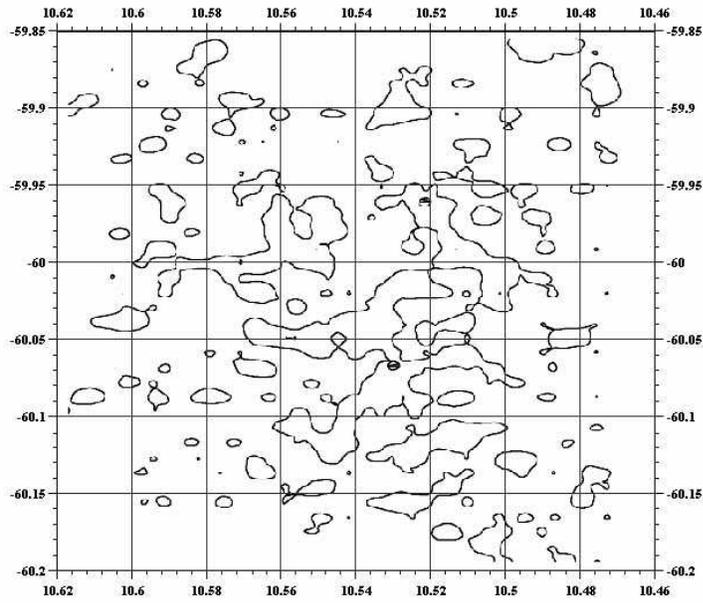}}
\caption{Contour map of the stellar density of the cluster Cr
223.}
\end{figure}

\begin{figure}
\centerline{\includegraphics[width=8cm,height=10cm]{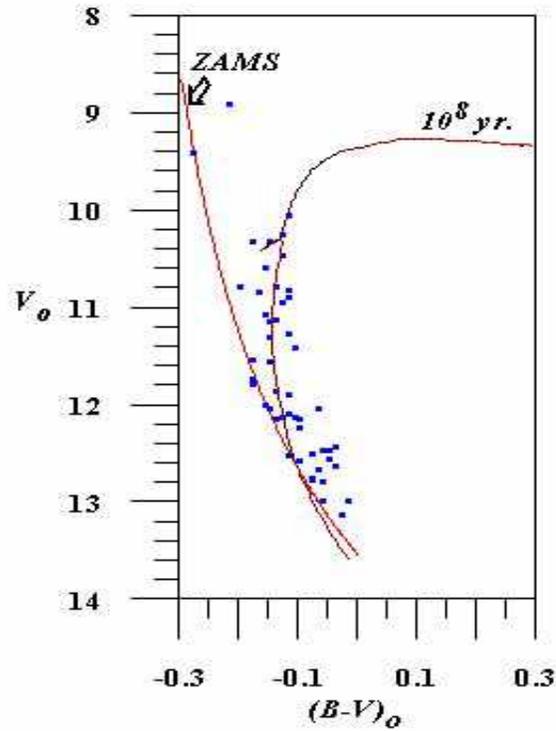}}
\caption{Age estimation of Cr 223 using the isochronous of Meynet
et al. \cite{MEY93}. The evolved sequence of the cluster implies
that it has an age of about 10$^{8}$ yr, or younger.}
\end{figure}

\begin{figure}
\centerline{\includegraphics[width=10cm]{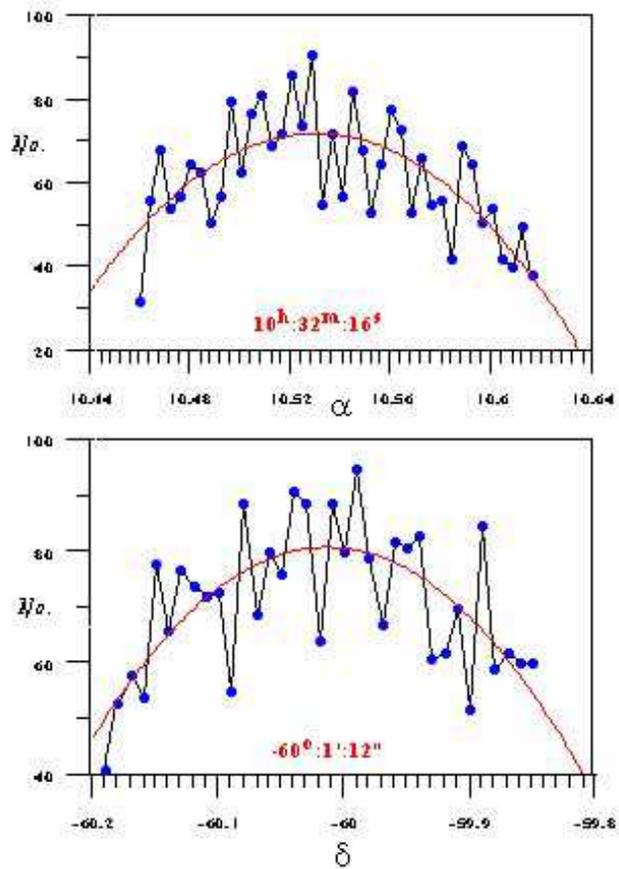}} \caption{The
center of symmetry about the peaks of $\alpha$ and $\delta$, which
have been taken to be the position of the cluster's center.}
\end{figure}

\begin{figure}
\centerline{\includegraphics[width=10cm]{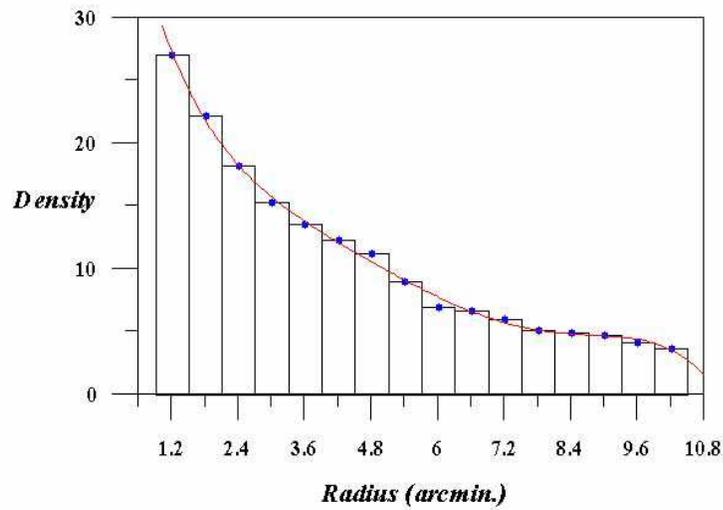}}
\caption{Radius determination of Cr 223; using the projected
density distribution. The radius of the cluster has been taken to
be 9 arcmin.}
\end{figure}

\begin{figure}
\centerline{\includegraphics[width=10cm]{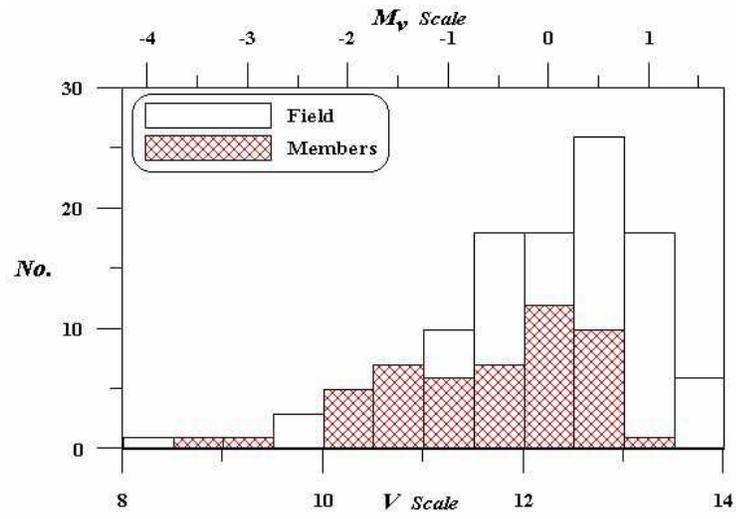}}
\caption{Luminosity function of Cr 223, the absolute magnitude
scale appears on the upper axis.}
\end{figure}

\begin{figure}
\centerline{\includegraphics[width=10cm]{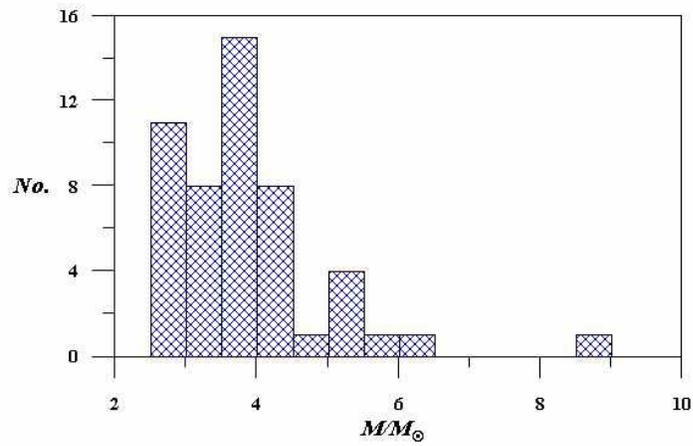}} \caption{The
mass function of Cr 223, most cluster's members have masses of
about 3.75 {\it M$_{\odot}$} }
\end{figure}

\newpage
\begin{table}
\caption{On the comparison with CL91.}
\begin{tabular}{lll}
\hline\noalign{\smallskip}Parameter&The present
work&CL91\\\hline\noalign{\smallskip}
Membership&49&46\\
E(B-V)&0.25 $\pm$ 0.015 mag.&0.26 $\pm$ 0.03 mag.\\
E(U-B)&0.18 $\pm$ 0.015 mag.&0.19 $\pm$ 0.03 mag.\\
(V-M$_{v}$)$_{o}$&12.25 $\pm$ 0.15 mag.&12.26 $\pm$ 0.20 mag.\\
Distance&2820$\pm$190 pc.&2830$\pm$260 pc.\\
$A_{v}$/E(B-V)&3.0&3.0\\
Z&-81 pc.&-96 pc.\\
Rgc&8.2 kpc.&--\\
X&-2.7 kpc.&--\\
Y&0.8 kpc.&--\\
Age&$\leq$ 10$^{8}$ yr.&3.6*10$^{7}$ yr.\\
Center&$\alpha$ = 10.5378$^{h}$& Near star no. 77\\
&$\delta$ = -60.02$^{\circ}$&\\
Diameter&$\approx$ 18 arcmin. ($\approx$ 14.8 pc.)&$\approx$ 15 arcmin.\\
Stellar density&See the text&--\\
Luminosity function&Peak lies at M$_{v}$=0.0 mag.&--\\
Total Luminosity&-4.4 mag.&--\\
Mass function&Peak lies at mass of 3.75 M$_{\odot}$&--\\
Total mass&$\approx$ 190 M$_{\odot}$&--\\
\hline{\smallskip}
\end{tabular}
\end{table}


\begin{thebibliography}{9999}
\bibitem[1986]{CL86}Clari$\acute{a}$, J.J. \& Lapasset, E., 1986, AJ, 91, 326.
\bibitem[1991]{CL91}Clari$\acute{a}$, J.J. \& Lapasset, E., 1991, PASP, 103, 998.
\bibitem[1931]{CO31}Collinder, P., 1931, Ann. Lund Obs. No. 2.
\bibitem[1988]{GCL88}Garcia, B., Clari$\acute{a}$, J.J. \& Levato, H., 1988, AP, \& SS 143, 377.
\bibitem[1965]{HO65}Hogg, A.A., 1965, Mem. Mt. Stromlo Obs. No. 17.
\bibitem[1953]{JOMO53}Johnson, H.L. \& Morgan, W.W., 1953, ApJ, 117, 313.
\bibitem[1981]{MER81}Mermilliod, J.-C., 1981, A\&A 7, 235.
\bibitem[1993]{MEY93}Meynet G., Mermilliod, J.-C. \& Maeder, A., 1993, A\&AS, 89, 477.
\bibitem[1982]{SCHM82}Schmidt-Kaler, Th.: 1982, Landolt-Bornstien \& H. H. Voigt, Numerical data and functional
relationships in Science and Technology, Group VI. 2, Subvol. b,
Springer - Verlag, Berlin.
\bibitem[2002]{TAD02}Tadross, A.L., Marie, M.A., Osman, A.I. \& Hassan, S.M., 2002, Ap \& SS, 282, 607.
\bibitem[1930]{TR30}Trumpler, R.J., 1930, Lick Obs. Bull., 14, 171.
\bibitem[1985]{VAN85}VandenBergh, D.A., 1985, ApJS, 58, 711.
\bibitem[1975]{WALL75}Wallenquist, A., 1975, Uppsala Astron. Obs. Ann., Band 5, no.8.
\end{thebibliography}
\end{document}